\documentclass{resonance}

\usepackage[english]{babel}

\usepackage{amsmath}
\usepackage{graphicx}
\usepackage[colorlinks=true, allcolors=blue]{hyperref}

\title{How to delay death and look further into the future if you fall into a black hole}
\author{A.V. Toporensky, S.B. Popov}

\begin{document}
\maketitle


\authorIntro{\includegraphics[width=2cm]{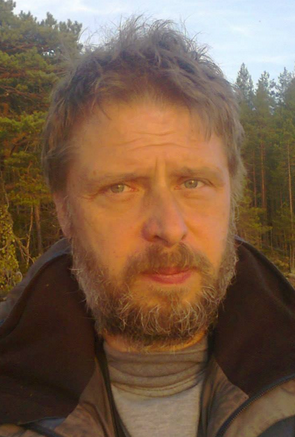}\\
Dr. Alexei Toporensky is working at the Sternberg Astronomical Institute and also is giving seminars at the HSE University, Moscow. His main research interests are in cosmology of the early Universe and theories of gravity. \\ \includegraphics[width=2cm]{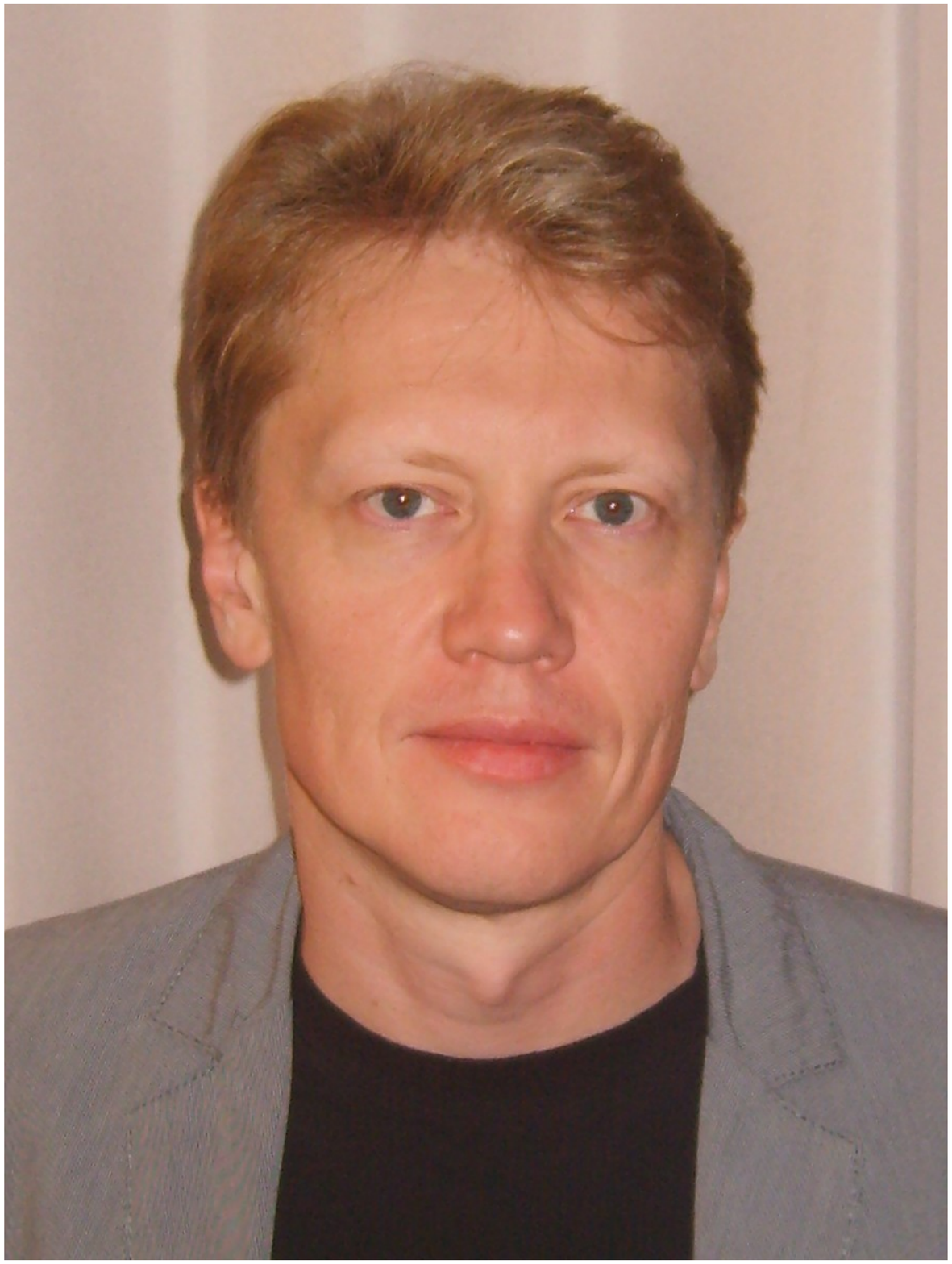}\\
Prof. Sergei Popov is working as the Sternberg Astronomical Institute and lecturing at the Lomonosov Moscow State University. Main scientific interests are related to evolution of compact objects, especially --- neutron stars. He also actively participates in popular science activity and public outreach.}

\begin{abstract}
 In this note, we present a pedagogical illustration of peculiar properties of motion in the vicinity and inside black holes. 
 We discuss how a momentary impulse can modify the lifetime of an object radially falling into a Schwarzschild black hole down to singularity. The well known upper limit for a proper time spent within a horizon, in fact, requires an infinitely powerful kick. 
 We calculate the proper time interval (perceived as personal lifetime of a falling observer) till the contact with the singularity, as well as the time interval in the Lema\^itre frame (which reflects how far into the future of the outer world  a falling observer can look), for different values of the kick received by the falling body. We discuss the ideal strategy to increase both time intervals by the engine with a finite power.   
 This example is suitable for university seminars for undergraduate students specializing in General Relativity and related astrophysical subjects. 
\end{abstract}

\artNature{GENERAL  ARTICLE}


\section{Introduction}

It is well known that even simple processes involving black holes (BHs) can produce apparent paradoxes. Many of them are discussed in the literature, others still wait for a detailed explanation. Analysis of such interesting problems which involve non-trivial aspects of General relativity (GR) effects can be very useful in a pedagogical practice. Bright examples of interesting phenomena with clear physical explanation can help students to understand BH properties and motivate them for further studies of GR. 

In this note we apply and develop some results from \cite{2020IJMPD..2930003T} to illustrate how a fall time into a Schwarzschild BH (till the singularity is reached) is modified by a momentary kick.
We used this example on seminars for masters-level students at the Lomonosov Moscow State University and HSE University. Discussions during these seminars demonstrated that the chosen topic helps to shed light on important aspects which are not well-understood by many students.  

The task is to derive and discuss necessary conditions to make the fall time as long as possible.
For a Schwarzschild black hole the ideal strategy for maximizing the proper
time can be easily derived from equations of motion. The metric in static coordinates $(t, r, \phi, \theta)$ has the form (we set the units so that $c=1$)
\begin{equation}
ds^{2}=-fdt^{2}+\frac{dr^{2}}{f}+r^{2}d\omega ^{2}\text{,}  \label{met}
\end{equation}%
where $f=1-r_g/r$, $r_g$ is the Schwarzschild radius $(r_g=2GM)$ and the angular part of the metric is  $d\omega ^{2}=d\theta ^{2}+d\phi ^{2}\sin ^{2}\theta $. 

We write equations of motion  for a particle in the space-time defined by eq.~(\ref{met}) in such a form that the proper time $\tau$ is present explicitly (the angular coordinates are chosen so that the motion to occur within the plane $\theta =\frac{\mathrm{\pi} }{2}$).
This form of the equations of motion can be obtained by using the existence of two integrals of motion: energy $E$ (since the metric is static)
and angular momentum $L$ (since the metric does not depend on $\phi$), see \cite{2020IJMPD..2930003T} for details.
\begin{equation}
\dot{r}=\pm Z\text{,}  \label{tz}
\end{equation}%
\begin{equation}
\dot{t}=\frac{\varepsilon }{f}\text{,}  \label{my}
\end{equation}%
\begin{equation}
m\dot{\phi}=\frac{L}{r^{2}}\text{.}  \label{L}
\end{equation}%
In the equations above 
a dot denotes differentiation with respect to the proper time $\tau$. We denote $\varepsilon = E/m$ and
\begin{equation}
Z=\sqrt{\varepsilon ^{2}-f\left(1+\frac{L^{2}}{m^{2}r^{2}}\right)}\text{.}
\label{zt}
\end{equation}%
 The $`+'$ sign in eq.~(2) indicates outward motion, while the $`-'$ sign corresponds to inward motion. Since we are interested in a motion under horizon we consider only the latter possibility in this paper.

From eq.~(\ref{tz}), it follows that a proper time required for a travel from the
horizon $r_g$ to $r_{1}<r_{g}$ is equal to%
\begin{equation}
\tau =\int_{r_{1}}^{r_g}\frac{dr}{Z}=-\int_{r_g}^{r_{1}}\frac{dr}{\sqrt{%
\varepsilon ^{2}-f\left(1+\frac{L^{2}}{m^{2}r^{2}}\right)}}.  \label{tau}
\end{equation}

For an object which is radially free falling into a black hole with zero velocity at infinity we have $\varepsilon=1$ and $L=0$,
so
the free-fall time from the Schwarzschild radius $r_g$ to singularity is:
\begin{equation}
    \tau_\text{ff}=-\int_{r_g}^{0}\frac{dr}{\sqrt{r_g/r}}=\frac23 r_g.
\end{equation}
Is it possible to increase this time for an object equipped with a rocket engine? The known answer is: `yes, but not much'. Since under the horizon both integrals of
motion enter equations in the quadratic form  with positive sign, the integral in eq.~(\ref{tau})  has its maximum value for $\varepsilon=0$, $L=0$. This maximum value is equal
to $(\mathrm{\pi}/2) r_g$ for $r_1=0$. So, the best strategy for maximizing the proper time under a horizon is to reach this optimal trajectory with two integrals of motion being equal to zero.
Note, that this trajectory is a geodesic one, since the motion  is two-dimensional and the two above mentioned integrals of motion represent a full set of integrals 
fixing a particular geodesic. 

The fact that the life time of an object with a jet engine is maximal at a geodesic trajectory (i.e., at a trajectory which requires the engine to be turned off after the ideal trajectory is reached) is in some sense a
counter-intuitive statement.  Sometimes it is claimed that this result easily follows from the fact that the interval $ds$ along a curve is just a proper time $d\tau$ of an observer following this curve, so as a geodesic  maximises $s$, it also maximises $\tau$. However, as it have been already discussed in 
\cite{2007PASA...24...46L} this is correct only if we consider a motion between two fixed space-time points. Only under this condition a geodesic is unique and maximises $\tau$. However, a singularity is not a space-time point! Its appearance differs for different coordinate systems, for the Gullstrand-Painlev\'e coordinates which we use below it is simply the line $r=0$. So that, we can not apply the above argument directly. Indeed, both $\varepsilon=1$ and $\varepsilon=0$ trajectories are geodesics, but the latter works obviously better. 

However, we will see soon that if the problem is considered in a bit more realistic situation --- with an engine of a finite power, --- the final
answer appears to be closer to intuitive feeling.

\section{Maximization of the proper fall time using a single-pulse engine}

 We consider a free-fall which is perturbed by a single engine thrust. 
It is assumed that this event momentary gives a peculiar velocity $V_p$ to the falling body. 
Afterwards, a free-fall continues till the object reaches the singularity. 

Velocity $V_p$ can be illustrated in the following simple way. The ship with an engine is free falling and just near it along the whole trajectory till the engine is turned on there is another free falling object.
$V_p$ is their relative velocity immediately after the momentary engine thrust.

 The thrust can be given just once. 
 We consider only situations when the thrusting  happens below the horizon ( for a more realistic modelling of a rocket behaviour near a BH horizon see, for example, \cite{2021arXiv211109240P}).  
 The task is to figure out the best strategy to maximize the time (in the falling body frame) till the singularity is reached for a given $V_p$. 
 We assume $\varepsilon =1$ which is approximately true for an object with its kinetic energy being small in comparison with $m c^2$ at a large distance from the BH.
 We also assume a radial fall. 
 
 Kinematics of the
 motion with respect to the frame with $\varepsilon=1$, $L=0$ (usually called the Lema\^itre frame) have been considered recently in detail \cite{2019JCAP...12..063T, 2021GrCo...27..126T}.
 The natural coordinate system associated with such a fall is the Lema\^itre one:
 \begin{equation}
 ds^2=-d\tilde {t}^{2}+(r_g/r) d\rho ^{2}+r^{2}(\rho ,\tilde{t})d\omega ^{2},
 \end{equation}
 with timelike coordinate $\tilde t$ and spacelike coordinate $\rho$,  so that the metric in these coordinates (in contrast to static coordinates $t$ and $r$) is regular at a horizon and everywhere else except for the singular point $r=0$. The Le\-ma\^i\-tre time is defined as $d\tilde t=dt+dr\sqrt{1-f}/f$ 
 and approximately coincides with the static time $t$ in the large distance limit where $f \sim 1$. On the contrary, its behavior is very different for small $r$, in particular it is finite at a horizon and inside a horizon.
 The former static coordinate $r$ 
 is now a function which for the Schwarzschild metric is known to be $r=r_g^{1/3} ((3/2)(\rho-\tilde{t}))^{2/3}$.
 This system is a synchronous one, and the  coordinate $\rho$ is constant during the free fall for particles with $\varepsilon=1$ and $L=0$. However, sometimes it is better to keep the coordinate $r$
 and use the metric in the following form (Gullstrand-Painlev\'e metric):
 \begin{equation}
     ds^2=-d\tilde{t}^{2}+(dr+vd\tilde{t})^{2}+r^{2}d\omega ^{2},
     \label{GP}
 \end{equation}
 with $v=\sqrt{1-f}$. 
 This form of metric is not so common in pedagogical literature, though the famous textbook \cite{2001ebhi.book.....T} is based mostly on it,
 where it is called as a `rain frame'.
 The price of using $r$ is the appearance of the off-diagonal term. The coefficient of this term is easily recognizable as a free fall velocity with respect to the stationary frame 
 $v=\sqrt{r_g/r}$.  Unlike a coordinate velocity, the velocity with respect to a frame has
 a direct physical meaning since it is by definition the velocity which is measured by a local observer
 belonging to the considered frame. In particular, this velocity is always subluminal.
 Under a horizon a stationary frame does not exist anymore, so $v$ looses its direct physical meaning becoming superluminal.  Note, that $v$ characterises the `rain frame' itself. A velocity of a particle with
 respect to this frame will be introduced in the next paragraph.
 
 The metric (\ref{GP}) has rather interesting properties. We mention here that the proper distance between two points at the same radius measured at $\tilde t=$~const hypersurface is simply the difference in $r$ coordinate \cite{1978PhRvD..17.2552G}. 
  Moreover, $\tilde t=$~const sections are flat. This  property is rather useful when we try to visualize the structure of a BH demonstrating together regions outside and inside the horizon  in a single picture. What is also useful,
 is that the rate of change of the radial coordinate $r$ of any object  with respect to the Lema\^itre time $\tilde t$  can be decomposed as
 \begin{equation}
     \frac{dr}{d \tilde t}=v-v_p,
 \end{equation}
 where $v_p$ is the velocity of an object with respect to the Lema\^itre frame.  So that, the second term in the right hand side has a local meaning, while two other terms  do not.  It is important, that this relation (which resembles the Galilean summation rule for classical velocities) is valid independently of the values of $v$ and $v_p$ everywhere, even under the horizon where $v$ exceeds the speed of light. Since the left hand side of this equation is a coordinate (i.e., not physical!) velocity, this does not contradict Special Relativity.
 The same situation is known for cosmology where the velocity of a distant object (which is a non-local entity, and, so, is not bounded from above) is decomposed into a sum of the Hubble and peculiar velocities (see, for
 example, an excellent discussion in \cite{2004PASA...21...97D}).
 To avoid confusion, we repeat that in our notation the value $V_p$ indicates the velocity caused by the engine, and, so, represents its power capability (i.e., it is a constant in a given set of conditions), while $v_p$ indicates a changing with time
velocity of any object with respect to the Lema\^itre frame and so can be considered as a variable peculiar velocity.

 We need also to take into account the fact that
 for a radially falling object with the
 velocity $v_p$ with respect to the Lema\^itre frame 
  its integral of motion can be written as:
 
 \begin{equation}
     \varepsilon =\frac{1-v v_{p}}{\sqrt{1-v_{p}^2}}. 
     \label {epsilon}
 \end{equation}
 The equation above also determines  the integral of motion  $\varepsilon$ of the spaceship after the engine have been used.
Both $v$ and $v_p$ should be taken at the point of the thrust. Since we assume that initially the spaceship 
have followed trajectory with $\varepsilon=1$, being at rest with  respect to the Lema\^itre frame, the value $v_p$
immediately after the thrust equals $V_p$. 

Now we return to the question of the proper time.
 We know that the proper time spent inside the horizon is maximized for $\varepsilon=0$. However, we see from the above equation that to reach such a trajectory is not an easy task! It can be done  only well inside a horizon since
 $\varepsilon=0$ corresponds to $v_p=1/v$, so that $r$ should be smaller than $r_g V_p^2$ to accomplish this maneuver. In particular, it can not be made exactly at a horizon since it requires an infinitely powerful engine ($\epsilon=0$ corresponds to $v_p=1$ at a horizon), so that the well known upper limit 
 for a proper time inside a horizon equal to $(\mathrm{\pi}/2) r_g$ is an unreachable limit, indeed.
 
 The natural goal for the captain of a spaceship would be to maximize the proper time since the fate of the ship is understood. 
 The captain can decide to wait till $r_c=r_g V_p^2$ is reached and then switch to the $\epsilon=0$ trajectory. The price needed to be paid is the time interval from the initial point (which we assume to be at the horizon) till the critical point which located  inside the horizon. During this period 
 the ship is falling  along the trajectory with $\epsilon=1$, i.e. far from the optimal one. Instead, the captain could decide to turn
 the engine on immediately, reaching the minimal possible value of $\varepsilon$ and then to fall along this still not optimized trajectory. Finally, the jet firing can be executed at some $r_\mathrm{on}$ in between $r_g$ and $r_g V_p^2$. To understand the best strategy in this setting we need to make corresponding numerical calculations.
 
 \begin{figure}
\centering
\includegraphics[width=0.9\textwidth]{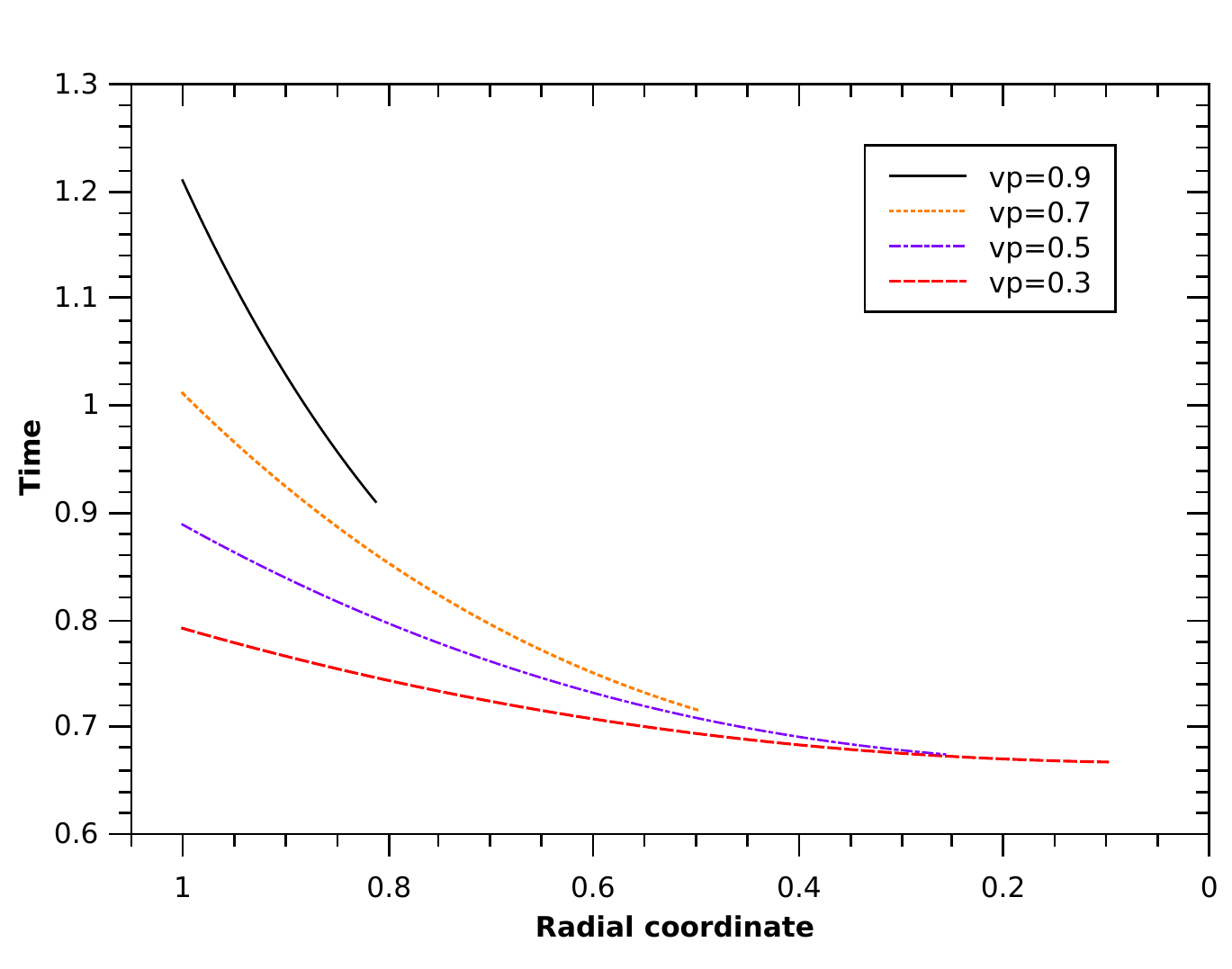}
\caption{\label{fig:rg}Falling from the horizon. A radial coordinate at the horizontal axis corresponds to the point where the engine is turned on. The vertical axis gives the time of falling from the horizon to the center of the BH. Different curves correspond to different values of $V_p$.}
\end{figure}
 
 We start calculations at the horizon. 
Then the time of the fall is written as a sum of two integrals:

\begin{equation}
    \tau= -\left(\int_{r_0}^{r_\text{on}} \frac{dr}{\sqrt{r_g/r}}+ 
    \int_{r_\text{on}}^0\frac{dr}{\sqrt{r_g/r + (\varepsilon^2-1)}}\right).
\label{time_f}
\end{equation}

 In this exercise we neglect tidal effects which would cause an actual death of an observer before reaching the singularity (for stellar mass BHs even before horizon crossing!). So, to be realistic we think about supermassive BHs where tidal effects close to the horizon are negligible. Of course, tidal distortion is always strong near the singularity, but it is easy to demonstrate, see e.g. \cite{2001ebhi.book.....T}, that effects are terribly strong just for a fraction of second, which is much shorter than the considered time of a free fall into a supermassive BH. Thus, we safely integrate down to $r=0$. 



We can see that the earlier the engine is activated --- the longer proper time is reached (see Fig.~1). So that, it is better not to wait until the optimal trajectory becomes available, but to react to the situation immediately. 
If the spaceship  captain recognizes the bad luck before the point $r=r_g V_p^2$, then the best strategy (i.e., switching on the engine for the full power as early as possible) can be considered as matching our `general' intuition.
Only if the critical point is already passed, the captain should deliberately switch off the engine at the trajectory with $\varepsilon=0$ despite some fuel is still unused.
 
In principle, the curves in Fig.~1 can be prolonged even further to the left, i.e. to the region which corresponds to a free fall from outside of the horizon. 
Our calculations show that curves there are still monotonic: the earlier is the kick given --- the larger is the proper time of a free fall. However, a radial kick outside the horizon is not the best option since the angular momentum enters eq.~(\ref{time_f}) with the negative sign for $r>r_g$ --- so, it is better to deflect the trajectory than to slow down the free fall. That is why we do not show this part of curves in our Fig.~1.

\section{Maximization of the Lema\^itre time}

Though  maximization of the proper time has its obvious motivation, this is not the unique reasonable strategy for a falling observer. The other option is to maximize the Lema\^itre time. The reason for 
this strategy is to observe as much as possible events in the `outer' world. Since the Lema\^itre time coincides with the static time $t$ for infinitely large $r$, the larger is the Lema\^itre time needed to reach the singularity --- the larger part of the history of the Universe surrounding the black hole in question can be witnessed.

Properties of the Lema\^itre free fall time are quite different from those considered in the previous section. To see this we can calculate the interval of $\tilde t$ for a radial trajectory with
$\varepsilon=0$. It is possible to obtain an analytical result using the fact that at such trajectory $v_p=-1/v$. Then the integral
\begin{equation}
    \tilde t=\int^{r_1}_{0} \frac{dr}{v-v_p}
    \label{Lt}
\end{equation}
which is giving us the free fall time from $r=r_1$ to singularity, can be expressed through $v$ only. After substitution $v=\sqrt{r_g/r}$ we get: 
\begin{equation}
    \tilde t=r_{g}\ln \left(\frac{{\sqrt{r_{g}}+\sqrt{r_1}}}{{\sqrt{r_{g}}-\sqrt{r_1}%
}}\right)-2\sqrt{r_{g}r_1}.
\end{equation}
The equation evidently diverges if $r_1 \to r_g$ --- there is no upper limit for the Lema\^itre free fall time! Again, this infinite limit is unreachable since it is impossible to switch to the trajectory with $\varepsilon=0$ 
exactly at the horizon. Note, however, that the ability to see a very remote future is limited only by the power of the observer's engine.

We can stop here for a moment and remind the reader that the ability to see infinitely remote future
during a free fall into a black hole is one of the most popular misconceptions in the black hole physics
which is often considered in teaching of GR (see, for example, \cite{2009PhyU...52..257G}). However, while crossing a horizon
along a geodesics can not help to witness a remote future, the powerful engine ignited at a horizon can help! An infinitely remote future still can not be seen, but the more power engine is available --- the longer is the period of history of the outer Universe which observed during the fall.

\begin{figure}
\centering
\includegraphics[width=0.9\textwidth]{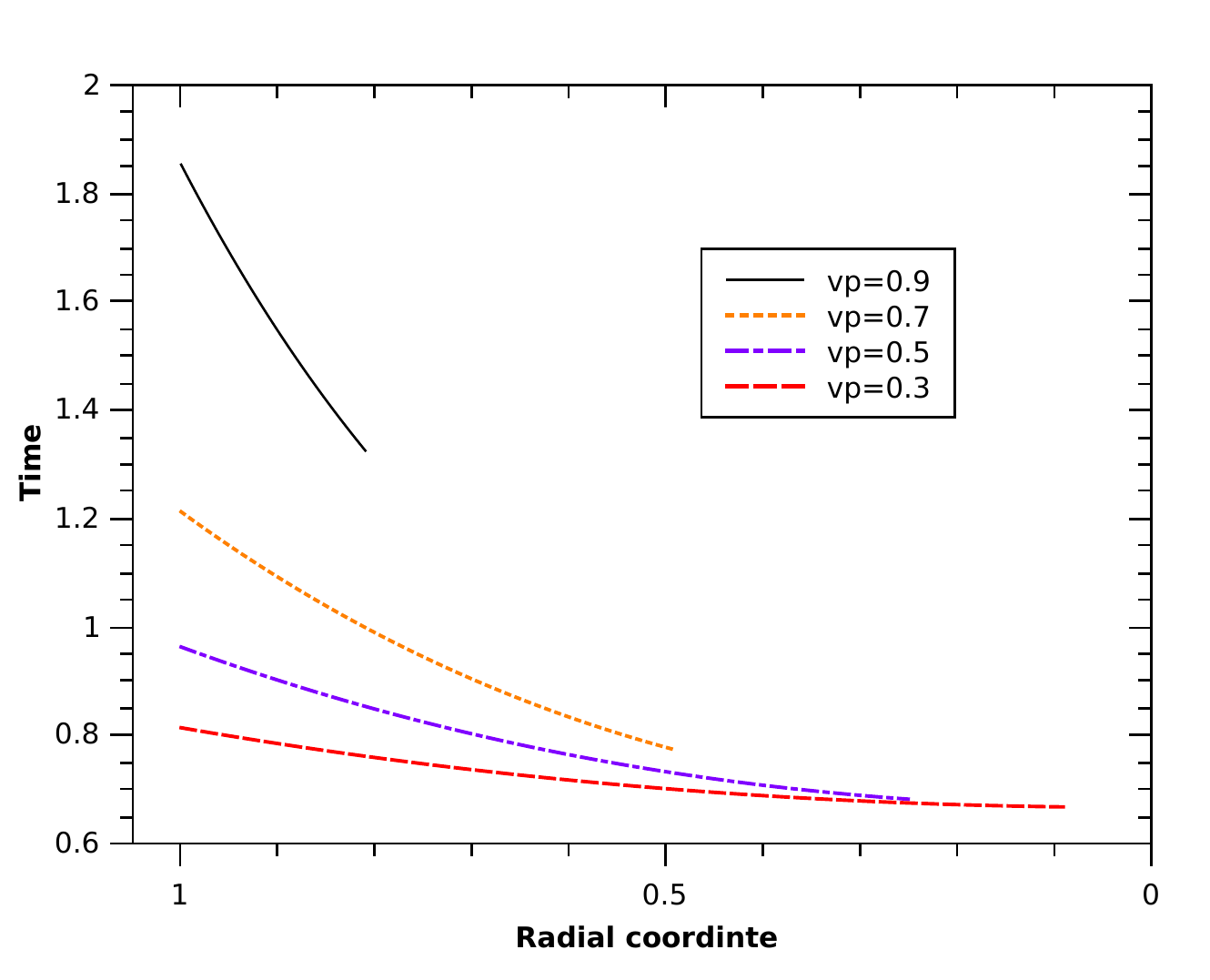}
\caption{\label{fig:lemetr} Fall time from the horizon in the Lemaitre frame. Different curves correspond to different values of $V_p$.  }
\end{figure}

Eq.~(\ref{Lt}) tells us also that the larger is $v_p$ --- the larger is the Lema\^itre free fall time, so the seemingly paradoxical situation when in some cases it is better to switch the engine off, never occurs while
maximizing $\tilde t$. The fact that the working engine can increase the coordinate time while decreasing the proper time have already been mentioned in \cite{2007PASA...24...46L} where the Eddington-Finkelstein time have been considered. The unreachable upper limit for the Lema\^itre time evidently corresponds to $v_p=1$, so it is given by the integral
\begin{equation}
    \tilde t_{max}=\int^{r_1}_{0} \frac{dr}{\sqrt{r_g/r}-1}.
\end{equation}
Note, that the maximum possible proper time is given by
\begin{equation}
    \tau_{max}=\int^{r_1}_{0} \frac{dr}{\sqrt{(r_g/r)-1}},
\end{equation}
so for any $r_1$ the maximum Lemaitre time is larger that the maximum proper time left to a singularity.
For a finite power engine with $V_p<1$ we need the evolution equation for $v_p$ which can be obtained by reversing eq.~(\ref{epsilon}):
\begin{equation}
v_p=\frac{v-\varepsilon Z}{v^2+\varepsilon^2}.
 \label{vp}
\end{equation}

Substituting eq.~(\ref{vp}) into eq.~(\ref{Lt}) we can obtain the Lemaitre time in the same way as we got the proper time in the previous section. The numerical results plotted in Fig.~2 show that in order to maximize $\tilde t$ we should
(as for maximizing $\tau$) switch on the engine the sooner the better. This again matches our intuitive feeling that it is better to start to struggle without any delay. Moreover, now this struggle should use all  available fuel independently of the observer's position. 

\section{Conclusions}

 Our teaching experience shows that problems like the one about maximization of the fall time serve as a good motivator for students. 
 However, during our seminars students never provided the standard (correct) answer to the question about the ideal strategy to maximize the proper time (`to reach the optimal trajectory and to turn-off the engine') before this problem was analyzed at a blackboard in details. Indeed, it can be hard to accept the counter-intuitive approach when the usage of the engine after some point might be considered as an act of sabotage. Luckily, the detailed analyzes (see above) shows that it is necessary to have an unrealistically powerful engine to reach the optimal trajectory close to the horizon. However, a realistic  strategy for a finite power engine is more intuitive and students typically provide such an answer: to use all the power as soon as possible. In this note we provided a detailed consideration, that this is the best approach when the optimal trajectory cannot be reached and the falling body did not pass the critical point ($r=v_p^2$), yet. 

In this note we also demonstrated that maximization of the proper and Lema\^itre time might be done following different procedures. If we already have $r<r_g V_p^2$ then the curiosity to know a remote  future (i.e., maximizing the Lema\^itre time) has its price in decreasing the proper life time. Note, however, that since the integral defining the Lema\^itre time diverges at a horizon, any maneuver deep inside the horizon can not increase it significantly. For example, consider a situation when a spaceship with the engine giving at most $V_p=0.9$ is located already at $r=0.49 r_g$. The proper time maximization strategy requires to use the boost only to $v_p=0.7$ which gives the proper time
equal to $0.275 (r_g/c)$. So, using the full power a hypothetical observer would sacrifice $0.024 (r_g/c)$ of his own proper time, instead increasing  the Lemaitre time for about $23 \%$ from $0.335 (r_g/c)$ to $0.413 (r_g/c)$.

The considerations of the present paper assumes that the initial kinetic energy of the spaceship is small with  respect to its rest energy, so we use the Lema\^itre frame to describe the motion. However, in our 
science-fiction set up we can easily imagine a situation when this condition is not satisfied. To deal with such a situation we need to use the kinematic with respect to a general radial free falling frame. We leave
this question for a future work.

\section*{Acknowledgements}
 We thank Dr. Oleg Zaslavsky from Karazin National University (Kharkiv, Ukraine) and Dr. Tarun Deep Saini from Indian Institute of Science (Bangalore, India) for discussions and useful comments. We also thank the referee for notes and suggestions which help to improve the paper.
 
 SP was supported by the Ministry of science and higher education of Russian Federation under the contract 075-15-2020-778 in the framework of the Large scientific projects program within the national project `Science'. 

\bibliographystyle{unsrt}
\bibliography{bh_fall}

\end{document}